# miniBELEN: a modular neutron counter for $(\alpha, n)$ reactions

*N* Mont-Geli[1*], *A* Tarifeño-Saldivia[2†], *L M* Fraile[3], *S* Viñals[4], *A* Perea[5], *M* Pallàs[1], *G* Cortés[1], *E* Nácher[2], *J L* Tain[2], *V* Alcayne[6], *A* Algora[2], *J* Balibrea-Correa[2], *J* Benito[3], *M J G* Borge[5], *J A* Briz[5], *F* Calviño[1], *D* Cano-Ott[6], *A* De Blas[1], *C* Domingo-Pardo[2], *B* Fernández[7,8], *R* Garcia[1], *G* García[4], *J* Gómez-Camacho[7,8], *E M* González-Romero[6], *C* Guerrero[7,8], *J* Lerendegui-Marco[2], *M* Llanos[3], *T* Martínez[6], *E* Mendoza[6], *J R* Murias[3], *S E A* Orrigo[2], *A* Pérez de Rada[6], *V* Pesudo[6], *J* Plaza[6], *J M* Quesada[7], *A* Sánchez[6], *V* Sánchez-Tembleque[3], *R* Santorelli[6], *O* Tengblad[5], *J M* Udías[3] and *D* Villamarín[6].

[1]Institut de Tècniques Energètiques (INTE), Universitat Politècnica de Catalunya (UPC), E-08028, Barcelona, Spain
[2]Instituto de Física Corpuscular (IFIC), CSIC – Univ. Valencia (UV), E-46071, Valencia, Spain
[3]Grupo de Física Nuclear (GFN) and IPARCOS, Universidad Complutense de Madrid (UCM), E-28040, Madrid, Spain
[4]Centro de Micro-Análisis de Materiales (CMAM), Universidad Autónoma de Madrid (UAM), E-28049, Madrid, Spain
[5]Instituto de Estructura de la Materia (IEM), CSIC, E-28049 Madrid, Spain
[6]Centro de Investigaciones Energéticas, Medioambientales y Tecnológicas (CIEMAT), E-28040, Madrid, Spain
[7]Departamento de Física Atómica, Molecular y Nuclear, Universidad de Sevilla (US), E-41012 Sevilla, Spain
[8]Centro Nacional de Aceleradores (CNA), Universidad de Sevilla (US) - J. Andalucía - CSIC, E-41092, Sevilla, Spain

**Abstract.** miniBELEN is a modular and transportable neutron moderated counter with a nearly flat neutron detection efficiency up to 10 MeV. Modularity implies that the moderator can be reassembled in different ways in order to obtain different types of response. The detector has been developed in the context of the Measurement of Alpha Neutron Yields (MANY) collaboration, which is a scientific effort aiming to carry out measurements of $(\alpha, n)$ production yields, reaction cross-sections and neutron energy spectra. In this work we present and discuss several configurations of the miniBELEN detector. The experimental validation of the efficiency calculations using $^{252}$Cf sources and the measurement of the $^{27}$Al$(\alpha,n)^{30}$P reaction is also presented.

## 1 Introduction

The production of neutrons through α-induced reactions plays an important role in many fields. Specifically, in nuclear astrophysics $(\alpha, n)$ reactions are a primary source of neutrons for the s-process [1] and a key mechanism in the synthesis of light r-process elements [2]. $(\alpha, n)$ reactions are also one of the main sources of the neutron-induced background in underground laboratories, which is a crucial issue in the low counting rate experiments carried out there [3,4]. Other fields of interest include nuclear technologies such as fission and fusion reactors [5,6] and non-destructive assays for non-proliferation and spent fuel management applications [7,8].

Experimental $(\alpha, n)$ cross-sections and production yields do exist in the literature. However, most of the experimental data was measured decades ago, is incomplete and/or presents large discrepancies not compatible with the declared uncertainties. Therefore, new measurements addressing the actual needs are required [9].

Most of the currently available data on $(\alpha, n)$ cross-sections and production yields has been obtained via activation measurements or direct neutron detection, which typically involves the use of neutron moderated counters. In the recent years several efforts have been carried out to develop new neutron moderated counters for $(\alpha, n)$ measurements [10,11].

In this work we present and discuss the design and the experimental characterization of miniBELEN, which is a modular and transportable neutron moderated counter developed in the context of the Measurement of Alpha Neutron Yields (MANY) collaboration, a coordinated effort aiming to carry out measurements of $(\alpha, n)$ production yields, reaction cross-sections, and neutron energy spectra. The α-beams are produced by the Spanish accelerator facilities at CMAM (Madrid) [12] and CNA (Sevilla) [13]. The detection setup includes the miniBELEN detector itself, the MONSTER array, which is a time-of-flight neutron spectrometer based on the use of BC501/EJ301 liquid scintillation modules [14], and a fast-timing array of LaBr$_3$(Ce) scintillation detectors of the FATIMA [15] type, which provides gamma detection with angular resolution capabilities.

## 2 Conceptual design of miniBELEN

Neutron moderated counters are neutron detectors based on the use of a low atomic number material to moderate the fast neutrons before being captured by thermal-neutron-sensitive proportional counters. In the case of detectors of the BELEN-type [16] such as miniBELEN, the moderator is made of High-Density PolyEthylene (HDPE) and $^3$He-filled detectors (1 inch diameter, 60 cm active length) are used as neutron counters.

---

* Corresponding author: nil.mont@upc.edu
† Corresponding author: atarisal@ific.uv.es

If the neutron energy spectrum is unknown, such as in most $(\alpha, n)$ reactions, the proper determination of the neutron yield requires a detection efficiency nearly independent from the initial neutron energy, namely a flat efficiency. This is typically achieved by optimizing the position of counters inside the moderator [11,17]. In miniBELEN, due to its modular structure, an alternative approach, the so-called composition method, has been used. The method is based on optimizing the contribution of each ring (i. e., a group of counters placed at the same distance from the centre of the detector) to the total detection efficiency. To that end the so-called composition functions are defined in the following way (assuming that the ring efficiencies can be considered independent magnitudes),

$$\varepsilon(E_n) = \sum_{i=1}^{n^\circ \, of \, rings} \varepsilon_{Ri}(E_n) \cdot f_i(E_n) \quad (1)$$

Being $\varepsilon(E_n)$ the total neutron detection efficiency, $\varepsilon_{Ri}(E_n)$ the detection efficiency of ring $i$, $E_n$ the original neutron energy and $f_i(E_n)$ the composition functions.

In miniBELEN the composition functions are physically implemented by partially covering the active region of the $^3$He detectors with cadmium filters (see Figure 1) in such way that neutrons of any energy that are moderated below 0.5 eV are absorbed. Therefore, the composition functions are nearly independent of the initial neutron energy.

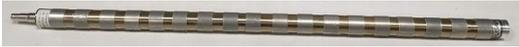

**Fig. 1.** 60-cm length and 1-inch diameter $^3$He-filled proportional counter partially covered with cadmium to provide the detector with a flat efficiency.

Modularity is the most interesting characteristic of miniBELEN. It means that the HDPE moderator is formed by several independent and smaller blocks which can be reassembled in different ways to obtain different types of response. Therefore, in practice, we have three detectors (configurations) for the prize of one.

### 2.1 miniBELEN configurations

miniBELEN consists of three different configurations using 10 or 12 cylindrical $^3$He-filled neutron counters manufactured by LND [18]. The gas pressure is generally 10 atm, although 4, 8 and 20 atm detectors are also used. Some counters are partially covered with 2 cm length and 0.5 mm thickness cylindrical cadmium filters implementing the composition functions listed in Table 1.

The basic components of the modular moderator are the 7x10x70 cm$^3$ blocks with a 1-inch central hole where the neutron counters are embedded. These blocks consist of seven smaller blocks (Figure 2b) which are assembled using two stainless-steel rods as shown in Figure 2a. Additional blocks are used to complete the whole detector.

**Table 1.** Composition functions ($f_i$) in miniBELEN.

| Configuration | Ring 1 | Ring 2 | Ring 3 |
|---|---|---|---|
| miniBELEN-10A | $f_1 = 0.4$ | $f_2 = 0.8$ | $f_3 = 1$ |
| miniBELEN-10B | $f_1 = 0.4$ | $f_2 = 0.875$ | $f_3 = 1$ |
| miniBELEN-12 | $f_1 = 0.325$ | $f_2 = 1$ | $f_3 = 1$ |

(a)      (b)

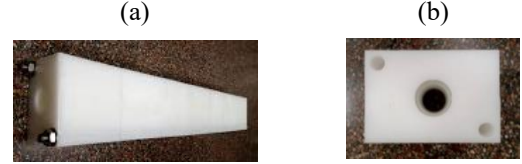

**Fig. 2.** 7x10x70 cm$^3$ HDPE blocks with a 1-inch central hole where the neutron counters are embedded (a). The blocks consist of seven 7x10x10 cm$^3$ smaller pieces (b) which are assembled using two stainless steel rods as shown in (a).

The HDPE moderator matrix can be divided in two parts: the "core", which is shown in blue in Figure 3, and the "reflectors", which are shown in red. The main objective of the "core" is to moderate neutrons while the primary objective of the 4 cm thickness "reflectors" is to back-scatter high-energy neutrons in order to increase the detection efficiency. The dimensions of the moderator are 58x43x70 cm$^3$ in configurations 10A and 12, while in miniBELEN-10B are 50x49x70 cm$^3$.

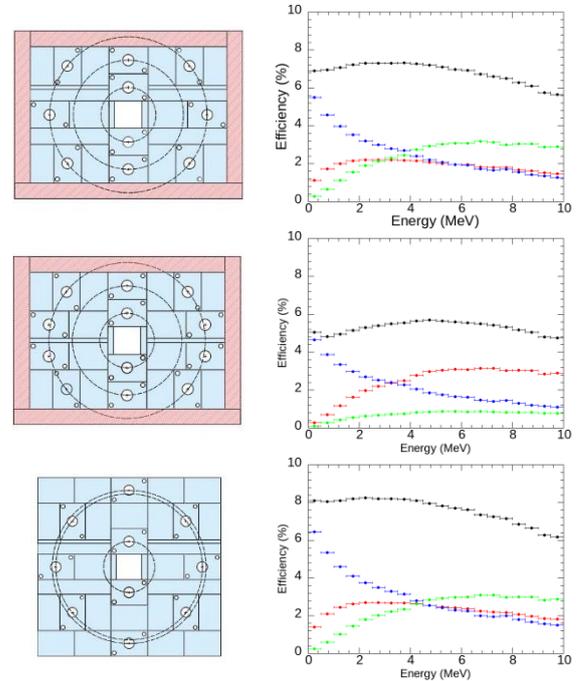

**Fig. 3.** From top to bottom: miniBELEN-10A, miniBELEN-12 and miniBELEN-10B. On the left, modular structure of the HDPE moderator. Blue: "core". Red: "reflectors" (see text). White circles: positions of the counters. Dotted lines: rings. White square: empty central region to place the neutron sources. On the right, neutron detection efficiency calculated using *Particle Counter* (see text). Red: ring 1. Green: ring 2. Blue: ring 3. Black: total. Bin size: 0.5 MeV. The y-axis error-bars are smaller than the size of the data-points.

## 2.2 miniBELEN efficiency

The neutron detection efficiency of each configuration of miniBELEN has been calculated using *Particle Counter* [19], a Monte Carlo application based on Geant 4 [20]. Calculations have been carried out using a point-like and isotropic neutron source placed at the centre of the detector. Results are presented in the form of a 0.5 MeV binned histograms in Figure 3.

Two figures of merit have been used for the optimization of the detector design, the nominal efficiency $\bar{\varepsilon}$ and the flatness parameter $F$.

$$\bar{\varepsilon} = \frac{1}{N_{bins}} \sum_{j=0}^{N_{bins}} \varepsilon_j \quad (2)$$

$$F = \frac{\varepsilon_M}{\varepsilon_m} \quad (3)$$

Both parameters are calculated from Monte Carlo simulations, being $N_{bins}$ the total number of energy bins, $\varepsilon_j$ the efficiency of bin $j$ and $\varepsilon_M$ and $\varepsilon_m$, respectively, the maximum and minimum values of the efficiency within a certain energy range. The values of these parameters for each optimal configuration are listed in Table 2.

**Table 2**. Nominal efficiency ($\bar{\varepsilon}$) and flatness parameter ($F$) of each configuration of miniBELEN. Only statistical uncertainties are reported.

| Neutron energy (MeV) | Conf. 10A | Conf. 10B | Conf. 12 |
|---|---|---|---|
| 0 - 5 | $\bar{\varepsilon}$ = 7.18(4) % $F$ = 1.062(8) | $\bar{\varepsilon}$ = 5.30(3) % $F$ = 1.18(1) | $\bar{\varepsilon}$ = 8.13(4) % $F$ = 1.038(7) |
| 0 - 8 | $\bar{\varepsilon}$ = 7.04(4) % $F$ = 1.125(9) | $\bar{\varepsilon}$ = 5.38(3) % $F$ = 1.18(1) | $\bar{\varepsilon}$ = 7.89(4) % $F$ = 1.152(8) |
| 0 - 10 | $\bar{\varepsilon}$ = 6.82(4) % $F$ = 1.30(1) | $\bar{\varepsilon}$ = 5.30(3) % $F$ = 1.19(1) | $\bar{\varepsilon}$ = 7.60(4) % $F$ = 1.33(1) |

miniBELEN-10A and miniBELEN-12 present a similar shape in the response and the most relevant difference between both configurations is the nominal efficiency. Nevertheless, it should be taken into account that, up to 5 MeV, the response of miniBELEN-12 is the flattest.

In the case of miniBELEN-10B, above 8 MeV its response is clearly flatter than in configurations 10A or 12. However, below 8 MeV the responses of configurations 10A and 12 are better in terms of the compromise between flatness and nominal efficiency.

## 3 Experimental validation of the detector response

The first validation involved the use of spontaneous-fission neutrons from $^{252}$Cf radioactive sources in order to determine the neutron detection efficiency of miniBELEN-10A. Due to the lack of a well-calibrated source, the Neutron Multiplicity Counting (NMC) technique has been used [21]. When comparing the experimental efficiencies to the calculated ones it has been found that *Particle Counter* tends to overestimate the neutron detection efficiency. Depending on the neutron source used, the overestimation ranges from 7 to 8%. This is not a surprising result since the same phenomenon appears in other works using similar setups [10,22]. It could be explained due to the experimental uncertainties of the neutron scattering cross-sections used in the code. For the calculations the $^{252}$Cf neutron energy spectrum from reference [23] has been used.

The second validation consisted on measuring the relatively well-known $^{27}$Al$(\alpha,n)^{30}$P production yields at CMAM using miniBELEN-10A. The target was a thick and high-purity natural aluminum foil and was placed at the center of the detector. A natural tantalum dummy target was used to characterize the neutron background. The $\alpha$-particles beam was accelerated at energies from 5 up to 8 MeV using the 5 MV tandem accelerator at CMAM and was passed through a natural tantalum collimator placed close to the target which also acted as a secondary electron suppression system. The beam current was measured by integrating the charge collected at the aluminum target.

When comparing the measured aluminum ($\alpha$,n) yields to previous measurements carried out by other groups [24–28], a good agreement is found. Relative discrepancies are no greater than 7% when the yields are obtained using the Monte Carlo calculated nominal efficiency. The systematic uncertainty in the neutron yields due to deviations of the efficiency from the perfect planarity is approximately 6%. The complete dataset from the experiment is expected to be published soon.

The third and last validation involved the neutron counting rates from $^{27}$Al$(\alpha,n)^{30}$P in each ring of miniBELEN-10A. These rates have been used to calculate the value of $Q_{ij}$, which is defined as,

$$Q_{ij} = \frac{R_i/R_j}{D_i/D_j} \quad (4)$$

Being $R_i$ and $R_j$ the counting rates in rings $i$ and $j$ and $D_i$ and $D_j$ the number of neutrons which should be detected in these rings if the calculated responses are the real ones,

$$D_i = \int \varepsilon_{Ri}(E_n) S(E_n) dE_n \quad (5)$$

$\varepsilon_{Ri}(E_n)$ is the neutron response for ring $i$ and $S(E_n)$ is the $^{27}$Al$(\alpha,n)^{30}$P neutron energy spectrum measured by Jacobs and Liskien [28] at 5.5 MeV.

Relying on the spectrum measured by Jacobs and Liskien and assuming that the measured rates are correct, the value of $Q_{ij}$ should be as close to 1 as close are the calculated ring responses to the real ones. From the experiment we obtain that $Q_{12} = 0.89(4)$, $Q_{13} = 0.91(4)$ and $Q_{23} = 1.03(2)$. Taking into account the fact that at low energies the spectrum from reference [28] presents large relative uncertainties (up to 50%), we can conclude that the response calculations are consistent with the measured rates.

## 4 Summary

We have presented the design of the miniBELEN detector, which is a modular and transportable neutron moderated counter consisting on three different configurations. The detector has been designed with a nearly flat response up to approximately 10 MeV in order to carry out measurements of $(\alpha, n)$ production yields and reaction cross-sections. For the conceptual design an innovative method based on the optimization of the contribution of each counter to the total efficiency by using thermal neutron absorbers (cadmium) has been used.

Monte Carlo calculations of the neutron detection efficiency have been experimentally validated for configuration miniBELEN-10A using $^{252}$Cf neutron sources and through the measurement of the well-known $^{27}$Al$(\alpha, n)^{30}$P production yield.

This work has been supported by the Spanish Ministerio de Economía y Competitividad under grants FPA2017-83946-C2-1 & C2-2 and PID2019-104714GB-C21 & C22, the Generalitat Valenciana Grant PROMETEO/2019/007, both cofounded by FEDER (EU), and the SANDA project funded under H2020-EURATOM-1.1 Grant No. 847552. The authors acknowledge the support from Centro de Microanálisis de Materiales (CMAM) - Universidad Autónoma de Madrid, for the beam time proposal (*Comissioning of neutron detector systems for $(\alpha, n)$ reaction measurements*) with code P01156, and its technical staff for their contribution to the operation of the accelerator.